\documentclass[aps, pra, twocolumn, superscriptaddress]{revtex4-2} 
\usepackage{amssymb, amsmath, amsthm}
\usepackage{xcolor}
\usepackage{graphicx}
\usepackage{hyperref}
\usepackage{comment}

\graphicspath{{Figures/}} 

\newcommand{\ket}[1]{| #1 \rangle}
\newcommand{\bra}[1]{\langle #1 |}

\newcommand{\mytitle}{
A dynamical theory for one-dimensional fermions with strong two-body losses:
universal non-Hermitian Zeno physics and spin-charge separation }

\begin{document}
 
\title{\mytitle}      

\author{Lorenzo Rosso}
\email{lorenzo.rosso@universite-paris-saclay.fr}
\affiliation{Universit\'e Paris-Saclay, CNRS, LPTMS, 91405 Orsay, France} 

\author{Alberto Biella}
\affiliation{Pitaevskii BEC Center, CNR-INO and Dipartimento di Fisica, Università di Trento, I-38123 Trento, Italy}
\affiliation{Universit\'e Paris-Saclay, CNRS, LPTMS, 91405 Orsay, France}
 
\author{Jacopo De Nardis}
\affiliation{Laboratoire de Physique Th\'eorique et Mod\'elisation, CNRS UMR 8089,
CY Cergy Paris Universit\'e, 95302 Cergy-Pontoise Cedex, France}

\author{Leonardo Mazza}
\affiliation{Universit\'e Paris-Saclay, CNRS, LPTMS, 91405 Orsay, France}

\begin{abstract}
We study an interacting one-dimensional gas of spin-1/2 fermions with two-body losses. 
The dynamical phase diagram that characterises the approach to the stationary state displays a wide quantum-Zeno region, identified by a peculiar behaviour of the lowest eigenvalues of the associated non-Hermitian Hamiltonian.
We characterise the universal dynamics of this Zeno regime using an approximation scheme based on a effective decoupling of charge and spin degrees of freedom, where the latter effectively evolve according to a non-Hermitian Heisenberg Hamiltonian. We present detailed results for the time evolution from initial states with one particle per site with either incoherent or antiferromagnetic spin order, showing how peculiar charge properties witnessed by the momentum distribution function build up in time. 
\end{abstract}

\maketitle

\section{Introduction}

Losses are ubiquitous in ultra-cold gases and in several situations their interplay with quantum physics is at the basis of a remarkable phenomenology. 
They can be used to detect quantum coherence and the onset of Bose-Einstein condensation~\cite{Kagan_1985, Burt_1997, Soeding_1999}, to stabilize quantum Hall states~\cite{Roncaglia_2010}, to cool the gas~\cite{Rauer_2016, Schemmer_2018, Dogra_2019}, or even to drive it through phases which violate the equilibrium thermodynamic Tan's relation~\cite{Bouchoule_2021}.

Several experiments have studied the dynamics of correlated one-dimensional quantum gases in the presence of two-body losses, both for bosons~\cite{Tolra_2004, Syassen_2008, Haller_2011, Franchi_2017, Bouganne_2017, Tomita_2017, Tomita_2019} and for fermions~\cite{Yan_2013, Zhu_2014, Sponselee_2018}. {The theoretical characterisation of the interplay between the unitary and lossy dynamics has thus emerged as an important challenge and has recently attracted attentiont~\cite{Baur_2010, GarciaRipoll_2009, Bouchoule_2020b, Rossini_2021, Bouchoule_2021BIS, Rosso_2022, Rosso2_2022, Orazio_2021}.} 
{In the presence of spinful gases,
the stationary states can be highly non-trivial as a consequence of spin conservation, and an incoherent mixture of entangled Dicke states is stabilised by losses, possibly useful for metrological purposes~\cite{Foss-Feig_2012, Nakagawa_2020, Nakagawa_2021, Rosso_PRA_2021}. However, despite various experiments with molecular~\cite{Yan_2013,Zhu_2014} and atomic gases~\cite{Sponselee_2018} have been performed to realised such incoherent mixture of entangled Dicke states, they were not able to certify the properties of the produced stationary states.}
Moreover, when losses are strong, the quantum Zeno (QZ) regime sets in, and a counter-intuitive increase of the gas lifetime takes place as the loss rate is augmented~\cite{Misra_1977, Itano_1990, Almut_2000, Beige_2000, Kempe_2001, Facchi_2001, Facchi_2002, Schuetzhold_2010, Stannigel_2014, Gong_2017, Froml_2019, Snizhko_2020, Biella_2021}. {As such, a many-body hard-core constraint takes place and atoms/molecules behave as fermionized (hard-core) fermions~\cite{Zhu_2014}, where
the losses are interpreted as fast and unread measurements.} 
Whereas a theory of the dynamics of spinless bosonic gases in this QZ regime has been developed~\cite{Rossini_2021, Rosso_2022}, the same is not true for spinful fermionic gases: this constitutes a fundamental hurdle both for the exploitation of the entangled stationary states and for the interpretation of existing experiments.

In this work we consider a spin-1/2 interacting fermionic gas in the presence of two-body losses. {Several theoretical works have addressed various aspects of the model and of its dynamics~\cite{Nakagawa_2020,Nakagawa_2021,Foss-Feig_2012, Yan_2013,Zhu_2014,Sponselee_2018, Rosso_PRA_2021}. In particular, Ref.~\cite{Rosso_PRA_2021} addressed the weakly-interacting and -dissipative regime, highlighting the impact of spin conservation on the full dynamics, going beyond determining the stationary properties. Here, we rather focus on the QZ regime, presenting a dynamical theory in such regime based on a spin-charge separation assumption.}
The numerical study of the associated non-Hermitian Hamiltonian and of the full master equation allows us to identify a QZ regime emerging when the system is strongly-dissipative or strongly-interacting. For a given initial state, a rescaling of times shows that the QZ dynamics is universal; we develop a simple and predictive theory for this, using the key assumption that spin-charge separation takes place, where the spin degrees of freedom are dissipatively cooled according to a slow non-Hermitian Heisenberg Hamiltonian, and with a cooling rate set by the charge correlations of the gas.
By considering two experimentally-relevant initial states, and by presenting new predictions on the dynamics of several experimental signatures, such as the density of the gas, its magnetic correlations, and its momentum distribution function, we aim to trigger a novel generation of quantitative experimental studies.

{The paper is organised as follows. In Sec.~\ref{Sec:model} we introduce the model and in Sec.~\ref{Sec:phase:diagram} we study its dynamical phase diagram. Then, in Sec.~\ref{Sec:QZ:HCF} we focus on the QZ regime characterised by hard-core fermionic particles, that we describe by means of spin-charge separation. In Sec.~\ref{Sec:RE-DSC} we present our dynamical theory for the dynamics in the QZ regime. Next, we compare the prediction of our theory to full quantum simulations regarding the density of the gas, its momentum distribution function and spin correlations (Sec.~\ref{Sec:comparison}). Furthermore, we also present a particular case of our general theory in Sec.~\ref{Sec:id:mf}, valid for the state with incoherent spin order. Finally, in Sec.~\ref{Sec:Concl} we draw our conclusions. }

\section{The model}
\label{Sec:model}
We consider a lossy gas of spin-1/2 fermions trapped in a one-dimensional (1D) optical lattice and prepared with one particle per site; several spin configurations will be considered. 
At time $\tau=0$ the optical lattice is lowered so that particles can tunnel to neighbouring minima; 
two-body losses can take place when two particles occupy the same site; our study can model experiments with fermionic molecules~~\cite{Yan_2013, Zhu_2014} or with atomic Ytterbium in a metastable excited state~\cite{Sponselee_2018}.
In Ref.~\cite{Foss-Feig_2012} it was shown that the stationary states of this loss process are incoherent mixtures of Dicke states.

\begin{figure}[t]
 \includegraphics[width=\columnwidth]{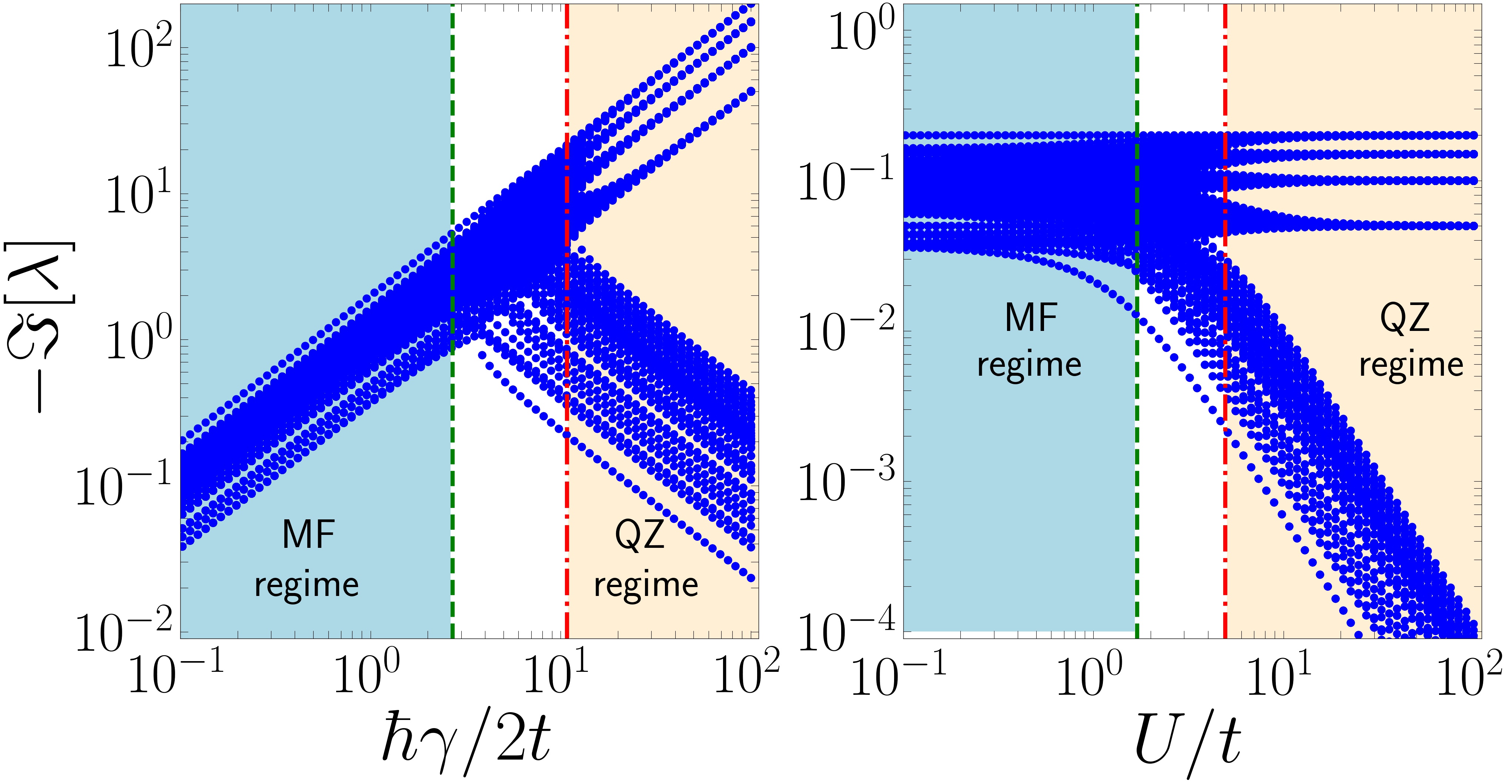}
 \caption{Imaginary part (with opposite sign) of the eigenvalues of the non-Hermitian Hamiltonian $\tilde H$ in the sector with $N = L = 8$ and $S_z = 0$ for $U = 0$ (left) and $\hbar\gamma = 0.1$ (right). {Green dashed vertical lines mark the boundaries of the MF region, while red dot-dashed lines mark the QZ one.} }
 \label{Fig:Imaginary:EVLS}
\end{figure}

Our goal is to characterise the dynamics of the gas, which is described by a Lindblad master equation for $\rho$, the density matrix of the system:
\begin{equation}
 \dot \rho = \mathcal L[\rho] = - \frac{i}{\hbar} [H, \rho]+ \sum_j L_j \rho L_j^\dagger - \frac 12 \{L_j^\dagger L_j, \rho \}.
 \label{Eq:Lindblad:MEQ}
\end{equation}
The 1D Hubbard Hamiltonian models the optical lattice in the single-band approximation: 
\begin{equation}
\label{Eq:Ham:Original}
H = -t \sum_{j,\sigma} \left( c_{j,\sigma}^\dagger c_{j+1,\sigma}+ H.c. \right)+ U \sum_j n_{j,\uparrow} n_{j,\downarrow};
\end{equation}
local two-body losses are described by the jump operators $L_j = \sqrt{\gamma} c_{j ,\uparrow} c_{j, \downarrow}$.
The $c_{j,\sigma}$ operators satisfy canonical anticommutation relations ($j$ labels the site and $\sigma$ the spin) and the density operator is $n_{j,\sigma} = c_{j,\sigma}^\dagger c_{j,\sigma}$; the hopping amplitude is $t$ and $U$ is the interaction parameter, $\gamma$ is the loss rate. 
The problem depends on two effective parameters: $\hbar \gamma/(2t)$ and $U/t$.
The spin of the gas is described by 
\begin{equation}
 \vec S = \sum_j \vec S_j =\frac{\hbar}{2}\sum_j \sum_{\tau, \tau'} c^\dagger_{j,\tau}
 \vec \sigma_{\tau \tau'}
 c_{j,\tau'},
\end{equation}
where $\vec \sigma$ are the three Pauli matrices.
The dynamics conserves the spin, as it can be deduced from the jump operators, that annihilate a spin singlet~\cite{Foss-Feig_2012}.

\section{Details about the phase diagram}
\label{Sec:phase:diagram}

The specific focus of this work is the study of the dynamics in Eq.~\eqref{Eq:Lindblad:MEQ} in the quantum-Zeno (QZ) regime.
In order to define it in simple terms, we consider the non-Hermitian Hamiltonian associated to the problem~\cite{Ashida_2020}, 
\begin{equation}\label{eq:HnonH1}
\tilde H = H - (i\hbar/2)\sum_j L_j^\dagger L_j.
\end{equation}
As $L^\dagger_j L_j = n_{j,\uparrow} n_{j,\downarrow}$, this corresponds to the Hubbard model with complex interaction  $\xi= U - i \hbar \gamma/2$.
The decay times of the system depend on the imaginary part of the eigenvalues of $\tilde H$, which we compute numerically using the package QuSpin~\citep{Quspin01, Quspin02}. The asymptotic lifetime of the gas is determined by those eigenvalues whose negative imaginary part is closest to zero. The analysis of the complex eigenvalues shows the existence of two well-defined regions, see for instance the plot in Fig.~\ref{Fig:Imaginary:EVLS}, which will be analyzed in the next two subsections.

\subsection{Mean-field universality}

\begin{figure}[t]
\includegraphics[width=\columnwidth] {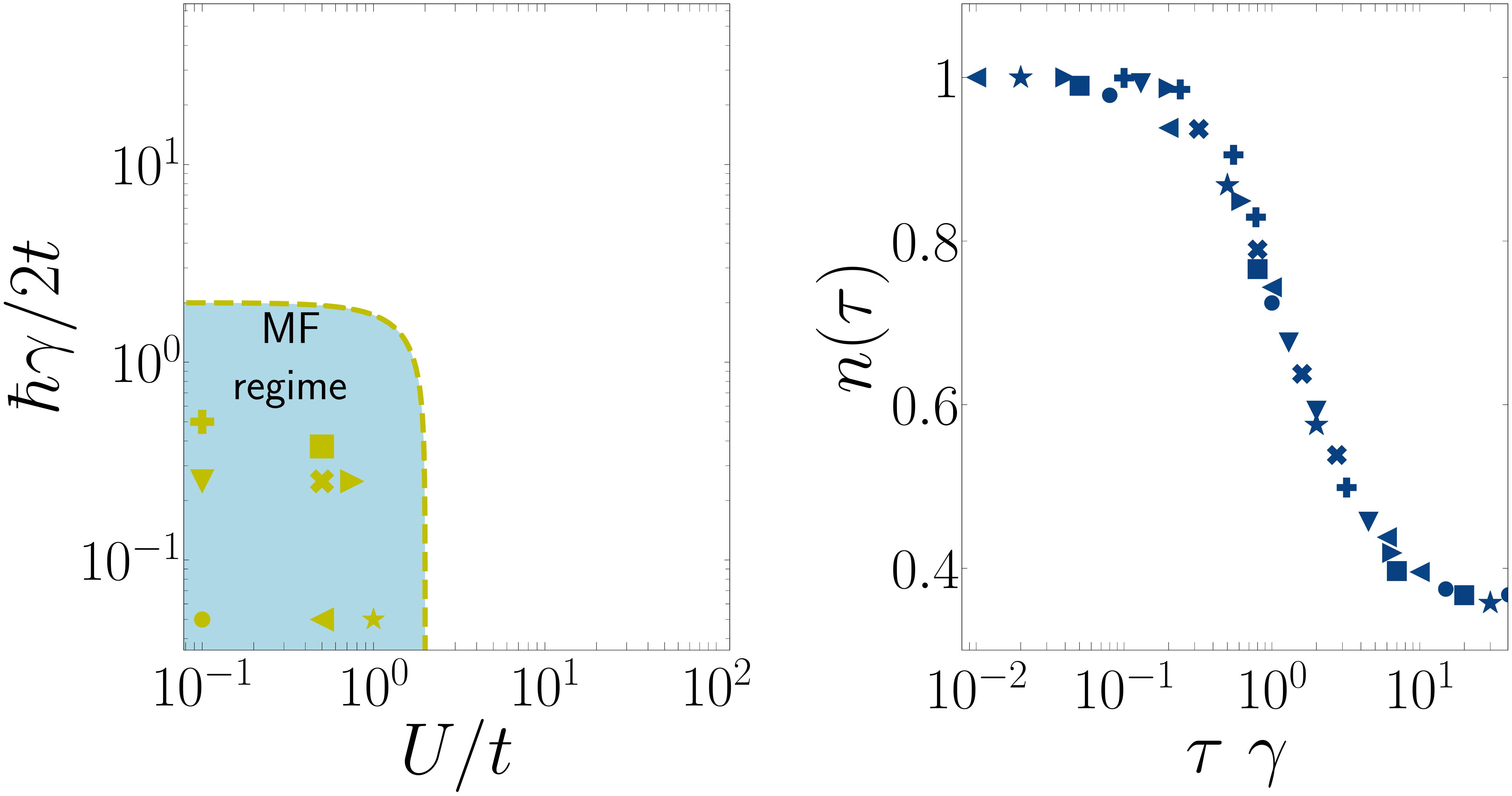}
 \caption{(Left) {The parameter space and the QZ region marked by $|\xi|/t>8.0$; the markers indicate the points of the phase diagram for which the Lindblad dynamics in Eq.~\eqref{Eq:Lindblad:MEQ} has been numerically} studied.
(Right) Universal QZ dynamics of the density of the gas for an initial N\'eel state according to the Lindblad equation~\eqref{Eq:Lindblad:MEQ}: different markers correspond to different points in the QZ region of the parameter space. The collapse is obtained by rescaling time with $\gamma$. Simulations are performed for $L=6$.}
 \label{Fig:evalue:mf}
\end{figure}

{In the weakly-dissipative and weakly-interacting regime appearing approximately for $|\xi|/t\lesssim 2$, the imaginary part of all eigenvalues increases linearly with $\gamma$, see Fig.~\ref{Fig:Imaginary:EVLS} (left panel).
The plot in Fig.~\ref{Fig:evalue:mf} shows the MF region and marks the boundary line $|\xi|/t = 2 $ with a dot-dashed line. In the right panel of Fig.~\ref{Fig:evalue:mf} we show the collapse of the curves by the appropriate rescaling with $\gamma$.}
 
\subsection{Quantum Zeno regime}

The QZ regime appears approximately for $|\xi|/t \gtrsim 8.0 $, as marked in Fig.~\ref{Fig:Imaginary:EVLS}, where there is a group of eigenvalues whose imaginary part decreases as $\hbar \gamma \times t^2 / |\xi|^{2}$ (see App.~\ref{App:derivation:me}), so that the lifetime of the gas increases with $\gamma$ \textit{or} with $U$. This result is more general than the standard
QZ effect because it takes place also for small $\gamma$.
In general, this increased lifetime follows from the absence of doubly-occupied lattice sites, so that losses are less effective. This can be the consequence of strong elastic interactions or of losses; in both cases, at long times, the gas is composed of long-lived \textit{hard-core} fermionic particles which never occupy the same site in pairs. 

\begin{figure}[t]
 \includegraphics[width=\columnwidth]{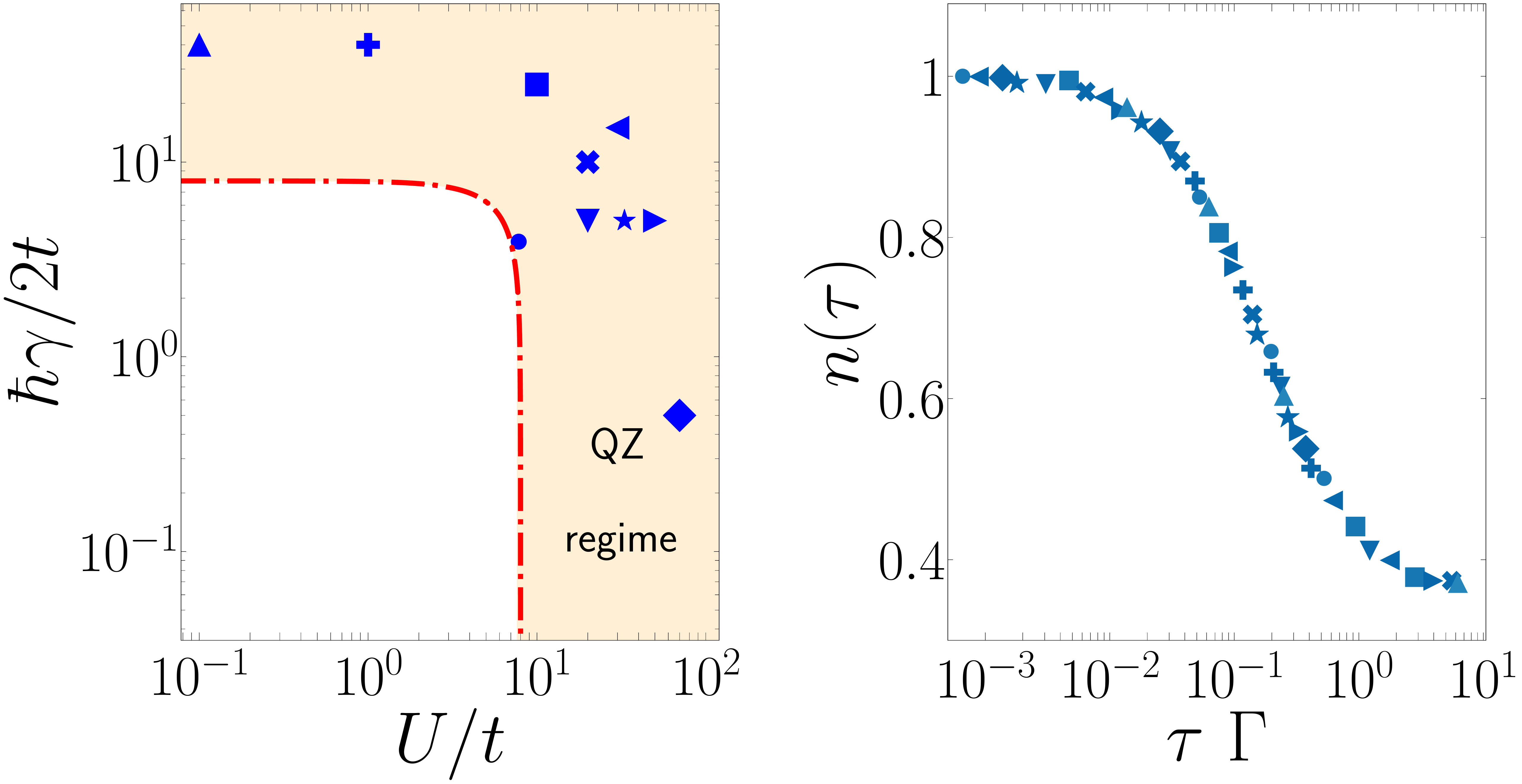}
\caption{(Left) {The parameter space and the QZ region marked by $|\xi|/t>8.0$; the markers indicate the points of the phase diagram for which the Lindblad dynamics in Eq.~\eqref{Eq:Lindblad:MEQ} has been numerically} studied.
(Right) Universal QZ dynamics of the density of the gas for an initial N\'eel state according to the Lindblad equation~\eqref{Eq:Lindblad:MEQ}: different markers correspond to different points in the QZ region of the parameter space. The collapse is obtained by rescaling time with $\Gamma$ in Eq.~\eqref{Eq:Zeno_decay}. Simulations are performed for $L=6$. }
 \label{Fig:RegimesTimeEvolution} \label{Fig:PhaseDiagram}
\end{figure}

The plot in Fig.~\ref{Fig:PhaseDiagram}(left)
shows the QZ region and marks the boundary line $|\xi|/t = 8.0 $ with a dot-dashed line.
We verify the existence of the QZ regime also in the master equation beyond non-Hermitian physics by performing a numerical simulation of the dynamics in Eq.~\eqref{Eq:Lindblad:MEQ} with the package QuTiP~\cite{Qutip01,Qutip02}. 
As a simple example, we initialize a N\'eel state with one particle per site and staggered spin structure, $\ket{\Psi_{\rm N}} = \ket{\ldots \uparrow \downarrow \uparrow \downarrow \uparrow \downarrow \ldots}$, and evolve it with parameters such that $|\xi|/t>8.0$:
the results are shown in Fig.~\ref{Fig:RegimesTimeEvolution}(right) and the densities $n(\tau)$ collapse by rescaling the time $\tau$ with the typical Zeno rate 
\begin{equation}
\Gamma = \gamma\times \frac{t^2}{|\xi|^2}. \label{Eq:Zeno_decay}
\end{equation}

{In between the MF and QZ regions lies an intermediate one which acts as a crossover. The latter is characterised by different behaviours and no appropriate rescaling of time to obtain the collapse of the dynamics has been found so far. A thorough investigation of the model in the whole phase space is particularly interesting, and could help understanding whether another transient behaviour is ``hidden'' between the MF and QZ regimes. We leave this study for future work. Now that the QZ regime has been properly defined, we focus on the dynamics of the gas in this limit.}

\section{Zeno limit and hard-core fermions }
\label{Sec:QZ:HCF}

The QZ regime is characterised by an extensive subspace of long-lived \textit{hard-core} states with at most one fermion per site. 
The effective master equation in this limit is given by~\cite{Foss-Feig_2012} :
\begin{equation}
 \dot \rho = \mathcal L'[\rho] = - \frac{i}{\hbar} [H', \rho]+ \sum_j L_j^{'} \rho L_j^{\dagger '} - \frac 12 \{L_j^{\dagger '} L_j^{'}, \rho \}.
 \label{Eq:Lindblad:MEQ:HCF}
\end{equation}

We introduce the hard-core fermion (HCF) operators $f_{j,\sigma}$ that satisfy all the properties of the $c_{j,\sigma}$ and in addition the constraint $f_{j, \uparrow} f_{j, \downarrow}=0$.
The Hamiltonian of the HCFs is the restriction of $H$ in Eq.~\eqref{Eq:Ham:Original} to the space of HCFs, and coincides with the Hubbard model in the $U \to \infty$ limit: $H' = - t \sum_{j,\sigma} ( f_{j,\sigma}^\dagger f_{j+1,\sigma}+ H.c. )$.
The jump operators follow from the judicious application of the dissipative Schrieffer-Wolff transformation~\cite{kato, kessler} to the original model and are accompanied by the appearance of the loss rate $\Gamma $ in Eq.~\eqref{Eq:Zeno_decay}, proper of the QZ regime, see App.~\ref{App:derivation:me}, 
\begin{equation}
L'_j = \sqrt{\Gamma/2} \sum_{\mu=\pm 1}\left( 
 f_{j,\uparrow} f_{j+\mu, \downarrow} - f_{j,\downarrow} f_{j+\mu,  \uparrow}\right) .
 \label{Eq:Lprime}
\end{equation}
The new jump operator annihilates spin singlets on neighbor sites $(j, j+1)$ and/or $(j-1,j)$.

In order to solve the HCF dynamics we decouple the spin and charge sectors, and propose an ansatz for the density matrix that is a product of the two:
\begin{equation}\label{eq:decoupling}
\rho(\tau) = \rho_{\rm c} (\tau)\otimes \rho_{\rm s} (\tau).
\end{equation}
This spin-charge separation is motivated by the well-known results for the Hamiltonian  in the $U\to \infty$ regime, that have been recently extended in Ref.~\cite{Nakagawa_2020} to the non-Hermitian Hamiltonian $\tilde H$. Whereas this decoupling does not automatically ensure that during the dynamics correlations between the two parts will not be created, we have observed a posteriori that this ansatz provides a good quantitative description of the dynamics.

Several methods have been proposed to deal with spin-charge separation at $U \to \infty$, which are taylored on the specific properties of HCFs~~\cite{Mielke1991b, Noga_1992, Ostlund_2006, Kumar_2008, Kumar_2009, Montorsi_2012, Nocera_2018, Tartaglia_2022}; we use here one that is due to Zvonarev et al. \cite{ZV}.
The Hilbert space of $N$ HCFs on a lattice of length $L$ is mapped to a one-dimensional model of $N$ spinless fermions on a lattice of length $L$ tensored with a spin-1/2 chain of length $N$.
For instance, the state with $L=4$ sites and $N=2$ particles $\ket{\uparrow \circ \circ \downarrow}$ is mapped onto the state $\ket{\bullet \circ \circ \bullet} \otimes \ket{\uparrow \downarrow}$.
The key point is that the spin chain carries information about the spin of each particle in an ordered way, from left to right and the HCF dynamics can swap the spin order only on the long time-scale of the super-exchange coupling and can be neglected in several situations.

We introduce the canonical spinless fermionic operators $a_j$ and the spin-1/2 operators $\vec \Sigma_j$ in order to describe the emerging charge and spin degrees of freedom, and reformulate the master equation in this novel language. The Hamiltonian $H'$ acts only on the charge and is just a free-fermion model, easily solved in momentum space:
$
 H' =  -2t \sum_k\cos k \; a_k^\dagger a_k.
$
The jump operator $L'_j$ removes two particles from neighbouring sites if they are in a spin singlet; in the new language it must take the form $L'_j = \Lambda_{j} a_j (a_{j+1}- a_{j-1})  $, where $\Lambda_{j}$ is an operator that checks whether the particles are in a spin singlet state and whose explicit expression is not necessary. 

Since the dynamics of the charges is much faster than the loss rate, as expressed by the inequality $t \gg \hbar \Gamma$,  we employ a time-dependent generalised Gibbs ensemble (GGE) approximation for the charge sector,(see Refs.~\cite{Lange_2018, Lange_2017, Lenarcic_2018, Mallaya_2019, Rossini_2021} for other examples of time-dependent GGE formalism applied to weakly-dissipative systems). 
The density matrix associated to the charge sector takes the form $\rho_{\mathrm{c}}(\tau) \sim \prod_k e^{- \beta_k(\tau) a_k^\dagger a_k}$, and it is fully determined by the occupation numbers $n_k(\tau)$, determined by the inverse temperatures $\beta_k(\tau)$, and given by $n_k(\tau) = \langle a^\dagger_k a_k \rangle_\tau $. The problem is now reduced to finding an evolution equation, i.e.~a rate equation, for the $n_k(\tau)$

\begin{figure}[t]
\centering
\includegraphics[scale=0.14]{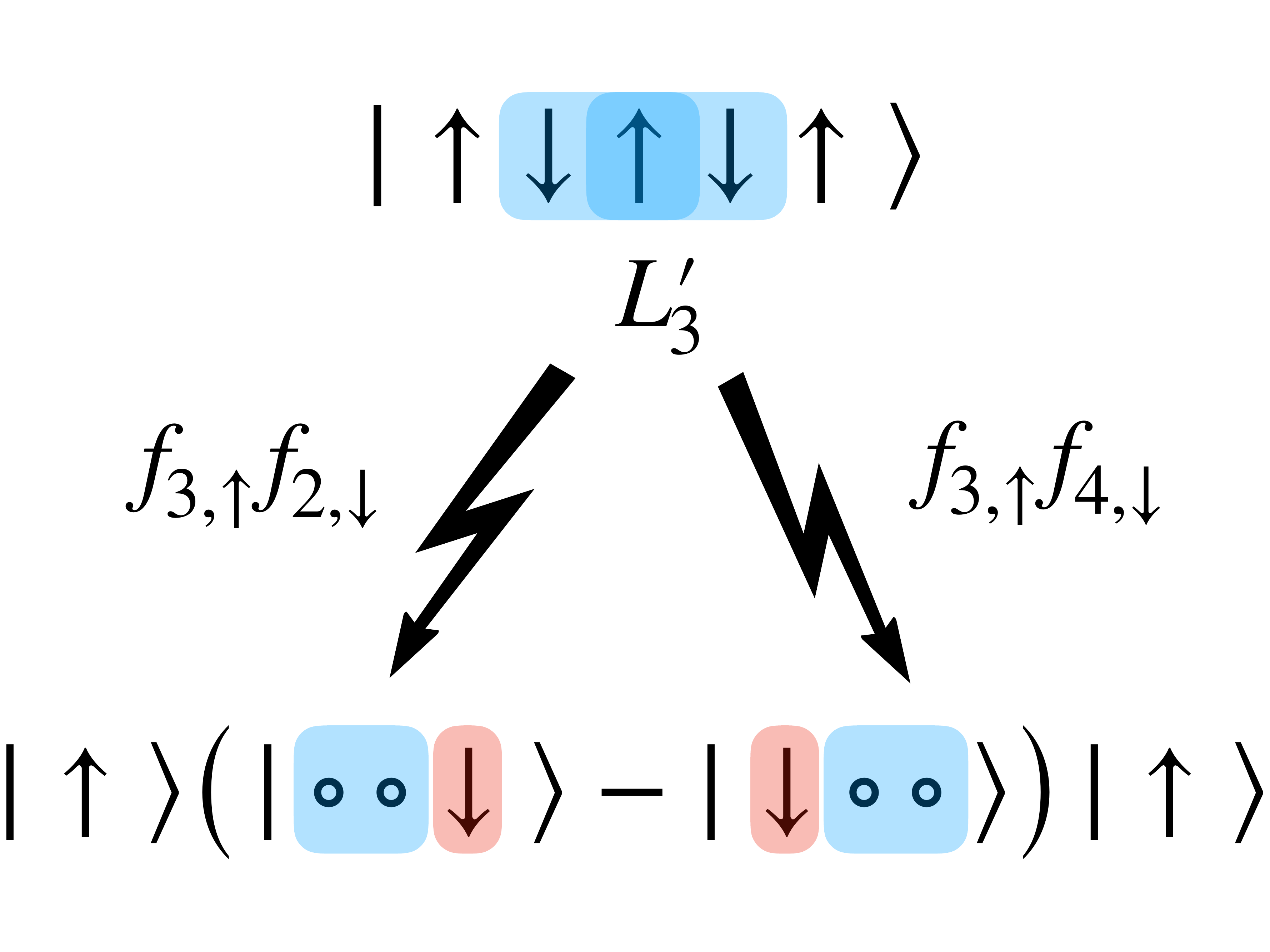} 
\caption{Sketch of a loss process taking place on $\ket{\Psi_{\rm N}}$ at site $j=3$. After the action of the jump operator $L_{3}^{'}$, the state of the system is proportional to that written at the bottom. In this state, space correlations are created only for the fermions with spin down (highlighted by red), implying the value $\delta_{\Psi_0}=1/2$ in \eqref{Eq:RateEqs:ConstantPi} for the antiferromagnetic state $\ket{\Psi_{\rm N}}$.
(Right) Dynamics of the number of particles $n(\tau)$ for an initial $\ket{\Psi_{\rm N}}$ state. The dashed curves for $L=10$ and $L=12$ are full simulations of the effective master equation for the \textit{hard-core} states~\eqref{Eq:Lindblad:MEQ:HCF}. The RE-DSC produces a very good description of the process, and the $\tau^{-1/3}$ is a fitting function.   }
 \label{Fig:g2}
\end{figure}

 \section{Rate-equation dissipative spin cooling}
 \label{Sec:RE-DSC}

The evolution of the occupation numbers $n_k(\tau)$ from the master eq. \eqref{Eq:Lindblad:MEQ:HCF} is given by:
\begin{equation}\label{eq:rate1}
\frac{ d}{d \tau} n_{k}(\tau) = \left \langle \sum_j L_j^{\prime \dagger} \left[a^\dagger_{k} a_k, L'_j \right] \right \rangle_{\tau}.
\end{equation}
An inspection of the jump operator in Eq.~\eqref{Eq:Lprime} elucidates that:
\textit{a)} losses do not simply take place when two particles get close by: it is also necessary that they are in a spin singlet; 
\textit{b)} the action of $L'_j$ on the state gives rise to quantum correlation only among a fraction of the whole (spinful) fermionic modes, according to the their initial spin structure. 
Let us first analyse point \textit{a)}: 
using the spin-charge decoupling we define the spin-singlet projection operator $\Pi_{j,j+1} =  (1-\frac{4}{\hbar^2}\vec \Sigma_j \cdot \vec \Sigma_{j+1})/4$ on two neighbouring sites of the spin chain, that represents the spins of the $j$-th and $j+1$-th particle and checks that they are in a spin singlet. In general, whether the two particles are neighboring depends on the charge part of the ansatz. Yet, we know that if they get close by, it is the operator $\Pi_{j, j+1}$ that checks if they are in a spin singlet. 
We define the density of spin singlets between consecutive fermions at time $\tau$, $\Pi(\tau) = \frac{1}{L}\sum_j\langle\Pi_{j,j+1} \rangle_\tau$, and we impose that the global rate of decaying of fermions is replaced as $\Gamma \to \Gamma  \, \Pi(\tau)$ for any momentum $k$.

Point \textit{b)} instead requires to notice that losses can create correlations in general only on a fraction $\delta_{\Psi_0}$ of fermions.
Let us consider for example an initial state with antiferromagnetic spin order, see Fig. \ref{Fig:g2}. In this case, after applying the operator $L'_3$, spin-down fermions develop spatial quantum correlations (one fermion with spin down, highlighted in red, is indeed delocalised over two sites), while fermions with spin up remain in a product state.
Since one loss process has created spatial correlations for half of the fermions, namely those with spin down, we have $\delta_{\Psi_{\rm N}} =1/2$, i.e. the factor $ \delta_{\Psi_{0}} $ takes into account the fraction of fermions among which spatial correlations are created after the action of the loss operator \eqref{Eq:Lprime}.

 {In order to understand why we need to use $\delta_{ \Psi_0} = 1/2$ for the N\'eel state, let us now consider an explicit calculation. We apply a jump operator to $|\uparrow \downarrow \uparrow \downarrow \uparrow \downarrow \rangle$; if we take
$j=3$,
the state is turned into a linear superposition:
\begin{equation}
 \frac{\ket{\uparrow \circ \circ \downarrow \uparrow \downarrow} - \ket{\uparrow \downarrow  \circ \circ \uparrow \downarrow}}{\sqrt 2};
 \label{Eq:Neel:Jump:Correlated}
\end{equation}
see also the sketch in Fig.~\ref{Fig:g2}. Now, it is easy to see that if we compute the momentum distribution function for spins $\uparrow$, $n_{k, \uparrow } = n_{\uparrow} = 1/3$. The other spin component, instead, features spatial correlations:
$n_{k,\downarrow} = n_{\downarrow}- \cos(2k)$.
In this case, the spins $\uparrow$ do not develop any spatial correlations, and for them a simple mean-field equation would be sufficient. On the other hand, the spins $\downarrow$ feature spatial correlation; of course, the situation would be reverted if the jump operator had acted on another site. This implies the use of $\delta_{ \Psi_{\rm N}}(\tau=0) = 1/2$.}

It is simple to generalise to generic spin states: $\delta_{\Psi_0}$ is given by the expectation value of the operator 
\begin{equation}
 \hat{\delta} = \frac{1}{2N} \sum_j [P^{\uparrow}_j  P^{\downarrow}_{j+1}   P^{\uparrow}_{j+2} +  P^{\downarrow}_j  P^{\uparrow}_{j+1}   P^{\downarrow}_{j+2}],
 \end{equation}
 with $P^{\uparrow}_j,P^{\downarrow}_j$ the spin up/down projector for the spin $\vec \Sigma_j$ of the $j$-th fermion.

 We shall therefore decompose the right hand side of Eq.~\eqref{eq:rate1} into a part proportional to $1-\delta_{\Psi_0}(\tau)$ where fermionic $k$ dependence is integrated away and a part proportional to $\delta_{\Psi_0}(\tau)$ where the full momentum structure of the fermionic expectation value (whose computation is reported in App.~\ref{App:solution:re}) is kept.

We are then in position to combine the two observations to 
obtain the following \textit{rate-equation} (for a full derivation see App.~\ref{App:deriv:re}):
\begin{align}
\frac{ d}{d \tau} n_{k}(\tau)  = - 4 \Gamma    \,\Pi(\tau) & \int_{-\pi}^{\pi} \frac{dq}{2 \pi} 
 \Big[ (1-\delta_{\Psi_0}(\tau)) + \nonumber \\ & + \delta_{\Psi_0}(\tau)
 \left(\cos k - \cos q \right)^2 \Big] n_q n_k.
 \label{Eq:RateEqs:ConstantPi}
\end{align}
 In order to close the equation and evaluate the time evolution of $\Pi(\tau)$ and $\delta_{\Psi_0}(\tau)$, we need to describe the spin dynamics. 
What happens to spin degrees of freedom is rather simple: the number of singlets decreases with time. This is exactly what is obtained once the spin-charge decoupling is applied to the non-Hermitian Hamiltonian in eq. \eqref{eq:HnonH1}: $H_s = -\Gamma/2 \sum_j \vec \Sigma_j \cdot \vec \Sigma_{j+1}$. Noting that $H_s = -\Gamma/2 \sum_j \left( 1 - \Pi_{j, j+1} \right)$~\cite{Nakagawa_2020,Nakagawa_2021, Yamamoto_2022}, we obtain a non-Hermitian time evolution~\cite{Sergi_2015} applied to an infinite spin chain without considering the fact that losses change the number of particles, since in the thermodynamic limit the number of particles, at any density, is infinite:
\begin{equation}
 \rho_{\mathrm{s}}(\tau) = \frac{e^{-\beta_s(\tau) H_s} \hspace{0.2cm} \rho_{\mathrm{s}}(0) \hspace{0.2cm} e^{-\beta_s(\tau) H_s} }{\mathrm{Tr} \left[ e^{-2 \beta_s(\tau) H_s} \hspace{0.2cm} \rho_s(0)  \right]}, 
 \label{Eq:rho:spin}
\end{equation} 
where $\rho_{\mathrm{s}} (0)$ is the density matrix describing the initial spin state. 
Note that $\rho_s$ depends on time only via $\beta_s$, and we  compute $\langle \Pi \rangle (\beta_s) = \text{tr}[\rho_s(\beta_s) \Pi]$ and $\delta_{\Psi_0} (\beta_s) = \text{tr}[\rho_s(\beta_s) \hat{\delta}]$ using an algorithm based on matrix-product-states \cite{Schollwck2011}.
In order to determine how $\beta_s$ flows with time, we observe that the spin is cooled each time a loss process takes place, and this depends on whether two particles are close-by, hence: $ \frac{d \beta_s}{d \tau}(\tau) = {\rm Tr}[\rho_{\rm c} (\tau)  \frac{1}{L} \sum_j  n_j n_{j+1}] $, which can be solved numerically together with Eq.~\eqref{Eq:RateEqs:ConstantPi}. The spin degrees of freedom are cooled down by this non-Hermitian evolution with a temperature that flows at a rate that depends on the charge properties of the gas, for this reason we dub Eq.~\eqref{Eq:RateEqs:ConstantPi} \textit{rate-equation dissipative spin cooling} (RE-DSC).

\section{Comparison with full quantum simulations}
\label{Sec:comparison}

We first  consider the system initialised in the N\'eel state $\ket{\Psi_{\rm N}}$.
We perform exact simulations of the Lindblad dynamics of eq.~\eqref{Eq:Lindblad:MEQ:HCF} for up to $L=12$ and compare it with our RE-DSC theory {with regards to three experimentally accesible quantities: particle density, nearest neighbors correlations and momentum distribution function.}

\begin{figure}[t]
\includegraphics[width=\columnwidth]{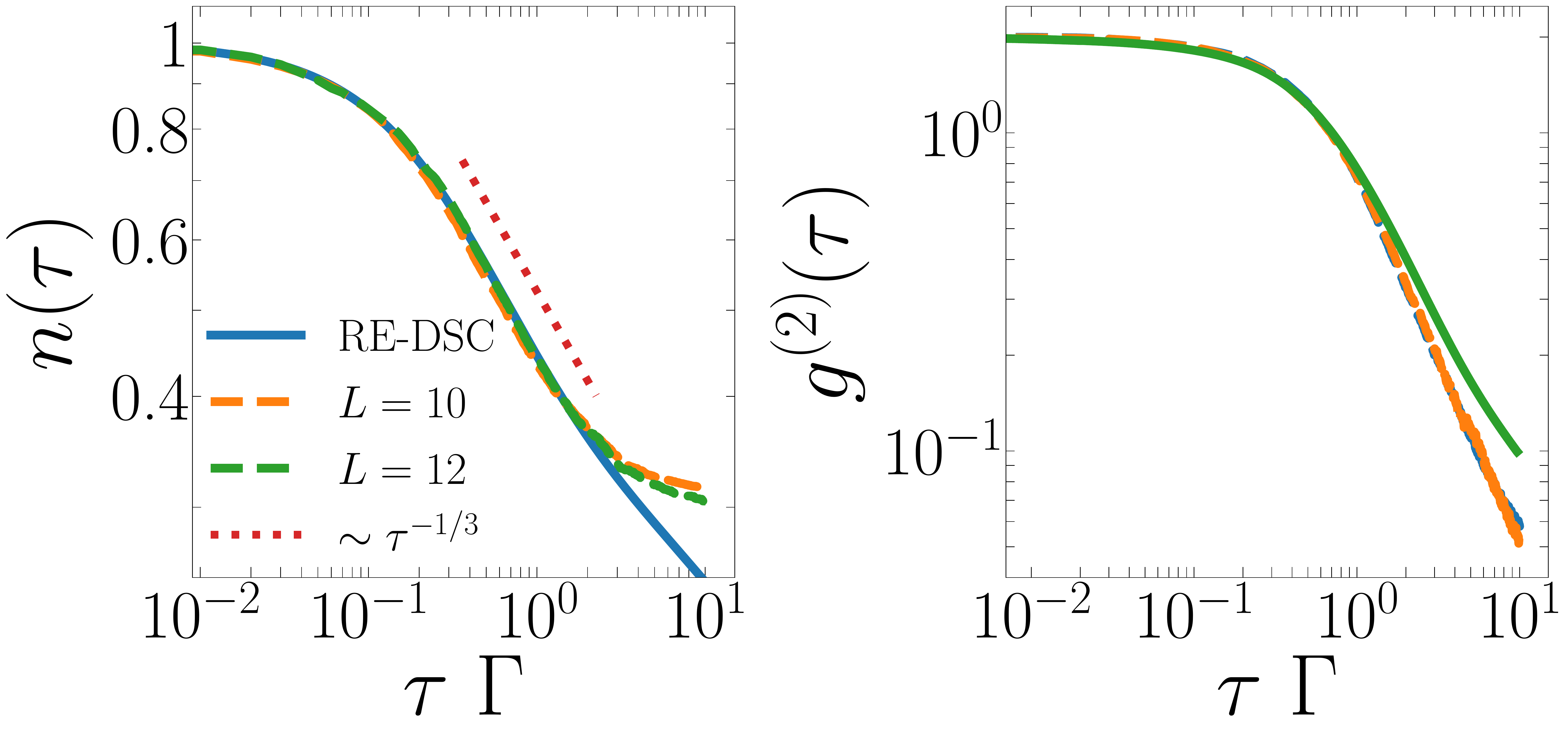}
 \caption{Dynamics of the number of particles $n(\tau)$ (left) and of the spin correlation function $g^{(2)}(\tau)$ (right) for an initial ordered N\'eel state. Dashed lines are obtained with numerical simulations of the master equation for $L=10$ and $L=12$. Solid green lines indicate the theoretical predictions obtained with our RE-DSC theory.}
 \label{Fig:data:neel}
\end{figure}

 {We show nn Fig.~\ref{Fig:data:neel} (left panel) the dynamics of the density of particles starting from a N\'eel state. We reproduce well the numerical data, and we fit a decay at \textit{intermediate} times compatible with the exponent $1/3$, i.e.  $n(\tau) \sim \tau^{-1/3}$. Let us notice that our theory addresses the thermodynamic limit of the model, and since the spin $\langle S^2 \rangle$ of the initial state scales as $L$ and not as $L^2$, we predict a final vanishing density $n(\tau \to \infty)= 0$, compatible with an algebraic decay~\cite{Rosso_PRA_2021}, as opposed to the finite density of the numerical simulations at finite sizes. The discrepancy at long time is fully under control.}
 
{Our theory gives full access to the correlations $g^{(2)} (\tau)$
\begin{equation}
g^{(2)}(\tau) = \frac{1}{L} \sum_i \left( \frac{\langle n_i n_{i+1} \rangle_\tau}{\langle n_i \rangle_\tau \langle n_{i+1} \rangle_\tau } - \frac{4}{\hbar^2}  \frac{\langle \vec S_i \cdot \vec S_{i+1} \rangle_\tau}{\langle n_i \rangle_\tau \langle n_{i+1} \rangle_\tau } \right),
\label{Eq:g2}
\end{equation} 
which in our spin-charge decoupling is proportional to the density of singlets $g^{(2)}(\tau) = 4 \Pi(\tau)$ since $\langle n_j n_{j+1} \rangle = n^2$ (within the t-GGE scheme) and $\frac{1}{\hbar^2}\langle \vec S_i \cdot \vec S_{i+1} \rangle = 1 - 4 \langle \Pi_{i, i+1} \rangle$ (by definition). We show in Fig.~\ref{Fig:data:neel} (right panel) the numerical data for $g^{(2)} (\tau)$ computed from the full numerical solution of the master equation starting from a N\'eel state, the data are in good agreement, up to finite-size effects, with our theoretical prediction obtained solving Eq.~\eqref{Eq:RateEqs:ConstantPi} with $\delta_{ \Psi_0} = 1/2$. The details about the numerical solution of Eq.~\eqref{Eq:RateEqs:ConstantPi} are presented in the App.~\ref{App:solution:re}}.

{In Ref.~\cite{Sponselee_2018}, the function $g^{(2)}(\tau)$ is fixed by a fitting formula which is then determined via experimental data; here we rather present a microscopic dynamical theory without fit parameters. 
Moreover, besides giving full access to the charge correlations, our theory predicts an algebraic decay.
Notice that this is in contrast with the exponential decay witnessed in numerical simulations of small system sizes.
As discussed in Ref.~\cite{Sponselee_2018}, an exponential decay of $g^{(2)}(\tau)$ is necessary to have some population in the stationary state.
Finally, the decay of $g^{(2)}(\tau)$ to zero for large $\tau$ indicates the creation of states whose spin-wavefunction is a Dicke state, which is one of the most intriguing aspects of this loss process~\cite{Foss-Feig_2012, Rosso_PRA_2021}}. 

\begin{figure}[t]
\includegraphics[width=\columnwidth]{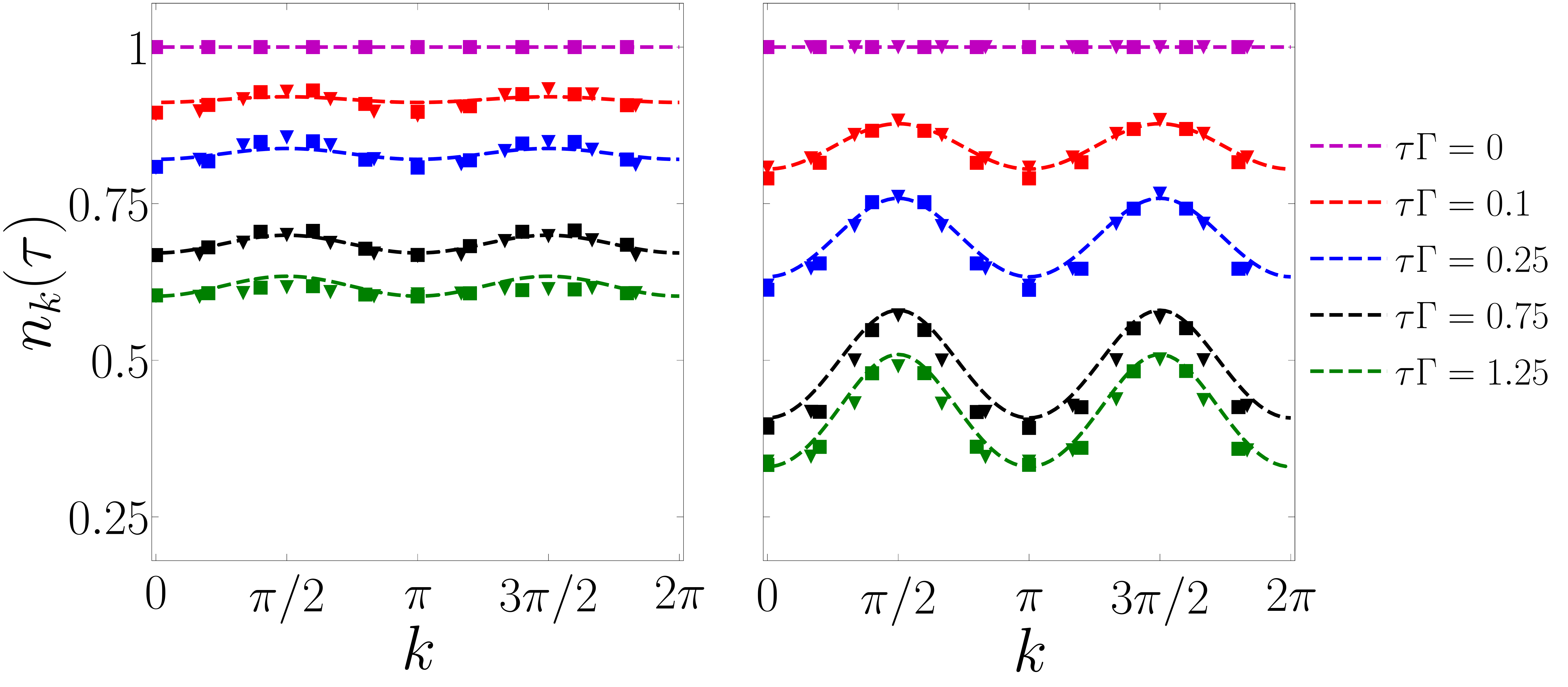}
\caption{Momentum distribution function $n_k(\tau)$ computed from the full numerical solution of the Lindblad master equation for $L=10$ (squares) and $L=12$ (triangles) at different times and the predictions of (RE-DSC)  eq. \eqref{Eq:RateEqs:ConstantPi} (dashed lines). Left: time evolution from the infinite temperature spin state. Right: time evolution from the N\'eel  spin state. }
 \label{Fig:nk}
\end{figure}

An important aspect of the RE-DSC theory is that it also allows to compute the momentum distribution function $n_k(\tau)$, that could be measured in an experiment. In Fig.~\ref{Fig:nk} we present a comparison of the numerical data with the results of our theory; the agreement is excellent and explains very well the appearance of two peaks at $k = \pm \pi/2$, which is  a distinctive feature of this Zeno regime of strong losses.

We also apply our theory to the initial state with exactly one particle per site and fully incoherent, infinite temperature $T = \infty$, spin structure. In this case we have $\delta_{T=\infty} \ll 1$ ({in fact $\delta_{T=\infty} = 1/8$,} see App.~\ref{App:delta:id}) at small and intermediate times, therefore we witness a much smaller modulation of the momentum distribution $n_k$ for this state compared to the N\'eel state,  Fig.~\ref{Fig:nk}, at these intermediate time scales, giving an excellent numerical confirmation of one of the major predictions of Eq. \eqref{Eq:RateEqs:ConstantPi}: i.e. initial states with large spin order leads to time-evolved states with strong inhomogeneities of $n_k(\tau)$ in k-space.

 We note that the t-GGE state $\rho_c(\tau)$ is a fermionic Gaussian state, and thus obeys Wick's theorem. Since $\langle a^\dagger_k a_q \rangle_{\tau} = \delta_{k,q} n_k(\tau)$, this gives direct acces also to charge correlation functions of the state. For instance, $\frac{1}{L}\sum_i \langle n_i n_{i+1} \rangle = \frac{1}{L^2} \sum_{k,q} \left ( 1 - \cos{(k-q)} \right) n_k n_q $, and, in particular, for both, left and right panel, momentum distribution functions in Fig.~\ref{Fig:nk} we obtain $\langle n_i n_{i+1} \rangle = n^2$. We expect this result to be true more in general when one considers initial states which are product states in the charge degrees of freedom.

\section{Spin identity initial state and mean field}
\label{Sec:id:mf}

\begin{figure}[t]
\includegraphics[width=\columnwidth]{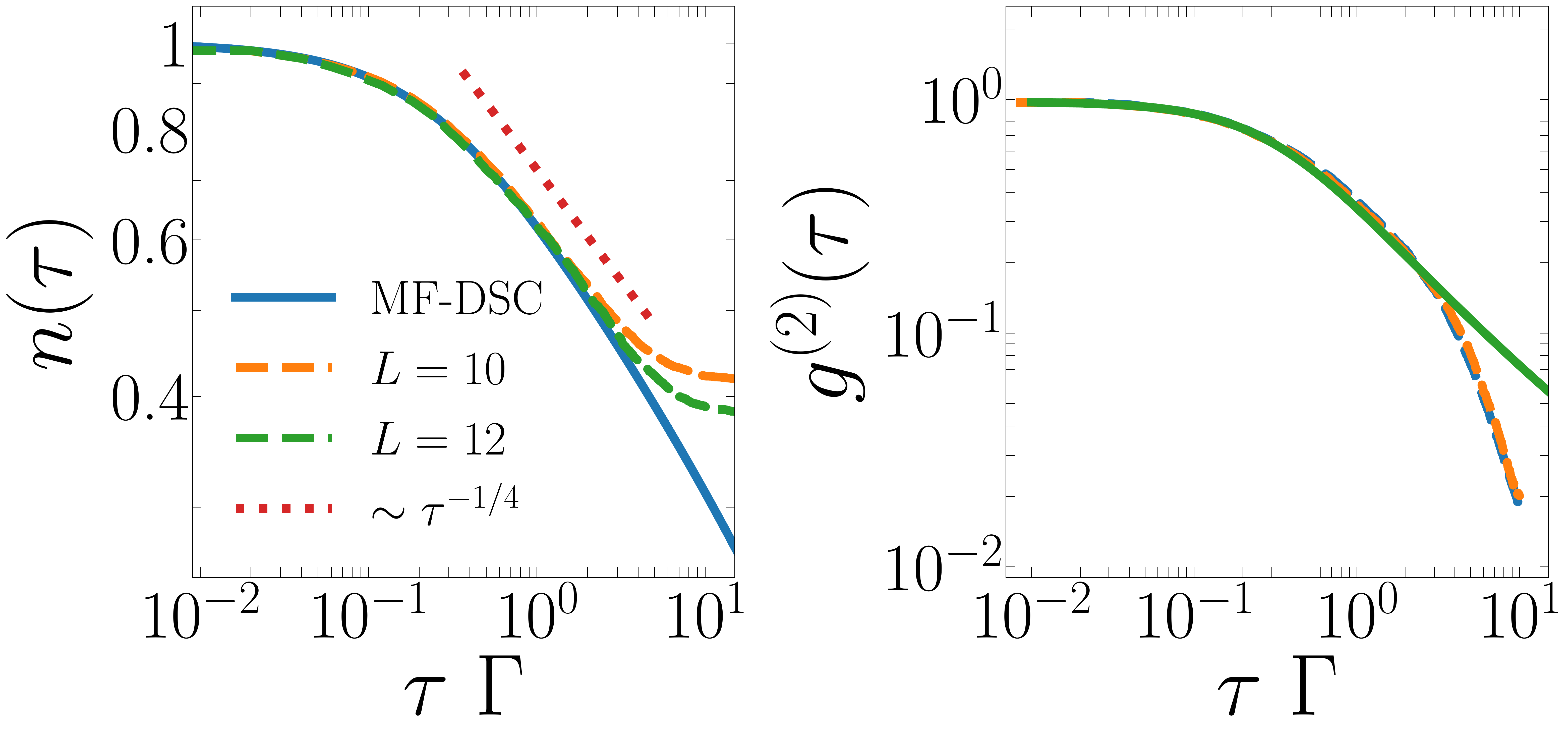}
 \caption{Dynamics of the number of particles $n(\tau)$ (left) and of the spin correlation function $g^{(2)}(\tau)$ (right) for an initial maximally-mixed spin state. Dashed lines are obtained with numerical simulations of the master equation for $L=10$ and $L=12$. Solid green lines indicate the theoretical predictions obtained with our MF-DSC theory.}
 \label{Fig:data:Id}
\end{figure}

{Our RE-DSC \eqref{Eq:RateEqs:ConstantPi} can therefore be view as an extension of the mean field equation first proposed in  Refs.~\cite{Baur_2010, Sponselee_2018}: $ \frac{d}{d \tau} n(\tau) = -   \Gamma  g_2(\tau) \; n(\tau)^2$, 
which is recovered by Eq.~\eqref{Eq:RateEqs:ConstantPi} under the simplification $\delta_{\Psi_0}=0$, which means that no spatial structure is created during the loss evolution. We now apply our theory to the initial state with exactly one particle per site and fully incoherent spin structure with $T =\infty$. As reported in App.~\ref{App:delta:id}, we have $\delta_{T = \infty}=1/8$. 
It is interesting to approximate
the latter to $\delta_{T = \infty} \sim 0$ so that the RE-DSC gives flat occupation of fermions in momentum space (no spatial correlation are created) and it can easily be manipulated into a mean-field-type of evolution for the total number of particles $n =  \int \frac{dk}{2 \pi} \ n_k$
\begin{equation}
 \frac{d}{d \tau} n(\tau) = -  4\, \Gamma \; \Pi\big(\beta_s(\tau)\big)  \; n(\tau)^2,
 \label{Eq:MF-DSC}
\end{equation}
where the spin temperature flows with time as
\begin{equation}
 \beta_s(\tau) = \int_0^\tau n^2 d\tau',
\end{equation}
as, in a state where $n_k(\tau) = n(\tau)$ there are no spatial correlations and thus $\text{tr}[ \rho_c(\tau)\frac{1}{L} \sum_j  n_j n_{j+1}] = n(\tau)^2$.
Knowing $\Pi(\beta_s)$, thus, a Runge-Kutta integration allows to compute easily $n(\tau)$. We conclude by displaying the data in Fig.~\ref{Fig:data:Id}, which show the evolution of $n(\tau)$ (left) and $g^{(2)} (\tau)$  (right) as a function of time. We reproduce well the numerical data which show a decay
at intermediate times compatible with the exponent $1/4$, i.e. $n(\tau)\sim \tau^{-1/4}$, after which numerical finite size effects become too relevant. Let us also mention here that, also for this initial state, we find an algebraic decay of correlations $g^{(2)}$ (Fig.~\ref{Fig:data:neel}) towards zero, indicating again the formation a Dicke-like spin wavefunction.}

\section{Conclusions}
\label{Sec:Concl}

In this work, we have presented a theoretical model for the quantum Zeno dynamics of a spin-1/2 gas in the presence of two-body losses. Our approach is based on a spin-charge decoupling that holds for strong losses or strong interactions, and it is benchmarked against numerical simulations of the full master equation for small systems up to $12$ lattice sites. Our model is based on the interplay between~(i) a non-Hermitian spin dynamics and~(ii) the build-up of a non-trivial momentum distribution function induced by losses; our study shows that it describes very well several observables at intermediate times. The main open point is whether it can describe the properties of the system at asymptotically-long times, and we leave this for future inspection. Our theory goes beyond previous studies by predicting the behaviour in real time of several observables: density, spin correlations and momentum distribution function; they can be tested in cold-atom experiments.

\textit{\textbf{Note added }---}
Recently, we became aware of an experimental article studying the dynamics of a SU(6) gas in presence of two-body losses~\cite{Honda_2022}. We believe that methods similar to those presented here could be used to describe that gas in the regime of strong losses. While completing the resubmission of this paper, we
became aware of a work discussing a different initial state for a $SU(N)$ dissipative gas~\cite{Yoshida_2022}. 

\section*{Acknowledgements}
We are grateful to M.~Zvonarev for sharing with us his work on the spin-charge separation in the one-dimensional Hubbard model.
Uncountable discussions with K.~Sponselee on the experimental setup are also gratefully acknowledged. We thank F. Essler for useful discussions on the Hubbard model. 
This work has been partially funded by  LabEx PALM (ANR-10-LABX-0039-PALM) and ERC Starting Grant 101042293 (HEPIQ).

\appendix
\section{Derivation of the effective master equation for hard-core fermions}
\label{App:derivation:me}

In this section we present some details about the derivation of the effective master equation governing the dynamics in the quantum Zeno regime
analyzed in the main text, that was called $\dot \rho = \mathcal L' [\rho]$.
We follow the method employed in Ref~\cite{GarciaRipoll_2009}; the final result has already been presented in Ref.~\cite{Zhu_2014} without derivation. 

We regroup the terms of the master equation 
\begin{equation}
 \frac{d \rho}{d \tau}=(\mathcal{V} +\mathcal{L}_{int}) \rho
\end{equation}
in the following manner:
\begin{equation}
    \mathcal{V} [\rho]= - \frac{i}{\hbar} [H_t, \rho], \quad \mathcal{L}_{int} [\rho]= - \frac{i}{\hbar} [H_{U}, \rho]+ \mathcal{D}[\rho]
\label{eq:fme}
\end{equation}
where the Hamiltonians are:
\begin{equation}
H_t = -t \sum_{j, \sigma} \left( c_{j, \sigma}^{\dagger} c_{j+1, \sigma} + H.c. \right), \quad H_U = U \sum_j n_{j, \uparrow} n_{j, \downarrow} 
\end{equation}
and the dissipation is:
\begin{equation}
\mathcal{D} [\rho] = \sum_j L_j \rho L_j^{\dagger} - \frac{1}{2} \left \{L_j^{\dagger} L_j, \rho \right \}, \quad L_j = \sqrt{\gamma} c_{j, \uparrow} c_{j, \downarrow}.
\end{equation}
This rewriting is useful to highlight the different orders of magnitude of the various term: $\mathcal{V}$ is a perturbation of order $t$, and we assume that in the quantum Zeno limit $|U- i \hbar \gamma /2| \gg t$.
In the following we are going to tackle the problem by means of a perturbative approach.

Let us start by focusing on the properties of $\mathcal L_{int}$, a non-Hermitian operator with infinitely many eigenstates. Exploiting a generalized version of Kato's method~\cite{kato} it is possible to expand: $\mathcal{L}_{int}= \sum_i \lambda_i \mathcal{P}_i$, using a complete set of projector operators with the following properties:
\begin{equation}
    \centering
    \mathcal{P}_i \mathcal{P}_j= \delta_{ij} \mathcal{P}_i, \hspace{0.2cm} \sum_i \mathcal{P}_i=1.
    \label{feq11}
\end{equation}
$\mathcal P_0$ projects the density-matrix over the hard-core fermion (HCF) subspace, which is stable. We call $\rho_0$ the density matrix restricted to the subspace.
With perturbative techniques,
it is now possible to construct the effective master equation governing the dynamics for the dominant term $\rho_0(t)$:
\begin{equation}
\frac{d}{d \tau} \rho_0 = \left(\mathcal L_1+ \mathcal L_2 \right) \rho_0  
\end{equation}
with
\begin{equation}
\mathcal L_1 = \mathcal{P}_{0} \mathcal{L}_{int} \mathcal{P}_{0}; \quad
\mathcal L_2 = \sum_{c } -\frac{1}{\lambda_c} \mathcal{P}_0 \mathcal{V} \mathcal{P}_c \mathcal{V} \mathcal{P}_0.
\end{equation}

\subsection*{First-order corrections: hard-core fermions}
Let us start by analyzing the first order corrections given by $\mathcal{L}_1$.
It can be shown~\cite{GarciaRipoll_2009} that $\mathcal{L}_1$ is equivalent to a Hamiltonian that has been projected within states without double occupancies. This is precisely a hard-core fermion gas under unitary Hamiltonian evolution $\mathcal L_1 [\rho_0 ] = -\frac{i}{\hbar} [H', \hspace{0.05cm}  \rho_0   ]$ where
\begin{equation}
     H' = -t \sum_{i=1} \big (f_{i+1 \sigma}^{\dagger} f_{i \sigma} + H.c. \big)
\end{equation}
where $f_{i \sigma}^{\dagger}$ and $f_{i \sigma}$ are the HCF operators satisfying the Clifford algebra plus the hard-core constraint.
The main result so far is that two body losses in the strongly dissipative regime lead to a coherent dynamics given by an hard-core fermion Hamiltonian.

\subsection*{Second-order corrections}
The second order Liouville can be recasted in a Lindblad form:
\begin{equation}
\mathcal{L}_2 [\rho_0] = \frac{i}{\hbar} \left [H'_2, \rho_0 \right] + \mathcal{D}_2 [\rho_0],
\end{equation}
where:
\begin{subequations}
\begin{align}
H'_2 &= -t_2 \sum_{j} L_j^{\prime \dagger }  L_{j}'; \\ 
\mathcal{D}_2 [\rho_0] &= \sum_j  L_{j}' \rho_0 L_j^{ \prime \dagger} -\frac{1}{2} \left \{ L_j^{ \prime \dagger}  L_{j}^{\prime}, \rho_0 \right \}.
\end{align}
\end{subequations}
The new set of jump operators describing the lossy dynamics is thus given by:
\begin{widetext}
\begin{equation}
    \centering
    L_{i}'= \sqrt{\frac{\Gamma}{2}}\left[  \left( f_{i, \uparrow} f_{i+1,\downarrow} - f_{i, \downarrow} f_{i+1,\uparrow} \right) + \left( f_{i, \uparrow} f_{i-1,\downarrow} - f_{i, \downarrow} f_{i-1,\uparrow} \right) \right],
\label{eq:jumps}
\end{equation}
with the coefficients given by:
\begin{equation}
t_2 = \frac{ 4 t^2 U}{\hbar^2 \gamma^2} \frac{1}{1 + \left (\frac{2 U}{\hbar \gamma} \right)^2} = \frac{U}{|\xi|} , \quad
\Gamma= \frac{ 4 t^2}{\hbar^2 \gamma} \frac{1}{1 + \left (\frac{2 U}{\hbar \gamma} \right)^2} = \frac{\gamma}{|\xi|},
\end{equation}
where we recall $\xi = U/t -i \hbar \gamma / 2t$ is the adimensional complex interaction defined in the main text.

Concluding, the effective master equation for the HCF has the following form (we dismiss here the notation $\rho_0$, that is not used in the main text):
\begin{equation}
\frac{ d}{d \tau} \rho (\tau) = -\frac{i}{\hbar} \left [ H'+H'_2, \rho(\tau) \right] + \sum_{j} \left[L'_{j} \rho(\tau) L_{j}^{\prime \dagger} - \frac{1}{2} \left \{ L_{j}^{\prime \dagger}  L'_{j}, \rho(\tau) \right\} \right].
\label{eq:eff_me}
\end{equation}
\end{widetext}
In the main text we do not make any explicit mention to $H_2'$ because it is completely irrelevant in the study that we carry out. This of course depends on the specific problem that we have chose to address, and this could not be the case for other situations.

\section{Derivation of the rate equations - Eq.~(12) of the main text}
\label{App:deriv:re}

In this section we consider the effective master equation for HCF that we have derived in Sec.~\ref{App:derivation:me}. 
Our goal is to present a derivation of the rate equations that appear in Eq.~(9) of the main text.
 
As we said in the main text, the Hamiltonian $H'$ has a simple form in the language of spin-charge separation:
\begin{equation}
 H' = -t \sum_j \left( a_j^\dagger a_{j+1}+ H.c.\right) = - 2 t \sum_k \cos k a^\dagger_k a_k.
\end{equation}
Similarly to what has been done for bosons in Ref.~\cite{Rossini_2021} we propose a generalised-Gibbs ensemble:
\begin{equation}
\rho_{\mathrm{c}}(\tau) = \prod_k \frac{e^{-\beta_k(\tau) a^\dagger_{k} a_k}}{\mathcal{Z}_k(\tau)} 
\label{Eq:tgge}
\end{equation}
fully determined by the occupation numbers $n_k(\tau)= \langle a^\dagger_k a_k \rangle_\tau$.
Using the master equation, we can take the time-derivative of
$n_{k}(t) = \mathrm {Tr} \left[ a^\dagger_{k} a_k \rho(t) \right]$, that reads:
\begin{widetext}
\begin{equation}
\frac{ d}{ d \tau} n_{k}(\tau) = \frac{i}{\hbar} \left \langle \left[ H' + H'_2, a^\dagger_{k} a_k \right] \right \rangle_t + \sum_j \left \langle L_j^{\prime \dagger} a^\dagger_{k} a_k L'_j - \frac{1}{2} \left\{ L_j^{\prime \dagger} L'_j, a^\dagger_{k} a_k \right\} \right \rangle_t.
\end{equation}
\end{widetext}

This expression can be simplified. First, we observe that $[ H', a^\dagger_{k} a_k ] = 0$; in fact, a more general relation holds,
$\langle \left[A, n_{k} \right] \rangle_{\tau} = 0$, that is valid for any operator $A$. Indeed:
\begin{equation}
\langle \left[A, n_{k} \right] \rangle_{\tau} = \mathrm{Tr} \left[ \rho(\tau) A n_{k} \right] - \mathrm{Tr} \left[ \rho(\tau) n_{k} A\right]
\end{equation}
but since $\left[n_{k}, \rho_{\mathrm{c}}(\tau) \right] = 0 $, using the cyclic property of the trace we obtain the result. The latter statement is true also for $A=H'_2$: neither of the two Hamiltonians influences the charge dynamics.

Focusing on dissipation,
we write:
\begin{equation}
- \sum_j \frac{1}{2} \left\{ L_j^{\prime \dagger} L'_j, a^\dagger_{k} a_k \right\} = - \sum_j L_j^{\prime \dagger} L'_j a^\dagger_{k} a_k + \frac{1}{2}  \left[ L_j^{\prime \dagger} L'_j, a^\dagger_{k} a_k \right]
\end{equation}
and we obtain:
\begin{equation}
\frac{ d}{d \tau} n_{k}(\tau) = \left \langle \sum_j L_j^{\prime \dagger} \left[a^\dagger_{k} a_k, L'_j \right] \right \rangle_{\tau}.
\end{equation}

In order to continue, we need to give an expression to $L'_j $ in the spin-charge language.
We propose the following one:
\begin{equation}
 L'_j = \sqrt{\frac{\Gamma}{2}} \Lambda_j a_j (a_{j-1}+ a_{j+1})
\end{equation}
where $\Lambda_j$ is a complicated non-local object acting both on spin and on charge that we are not able to treat exactly. 
The role of $\Lambda_j$ is to check that not only two particles come close by, but that they are also in a spin-spinglet: this condition is necessary for a loss event to occur.

At this stage, we find impossible to continue our work in an exact way. To begin with, we discuss what happens if we perform two ``reasonable'' approximations: first, that $\Lambda_j$ acts only on spin degrees of freedom, and second, that it simply checks whether two neighboring particles are in a spin-singlet channel. In particular, since we cannot say which one is the particle that is annihilated at position $j$, we will assume that it measures the average number of spin-singlets in the system:
\begin{equation}
 \Lambda_j^\dagger \Lambda_j \simeq \frac{1}{N} \sum_{\ell}\Pi_{\ell,\ell+1}, \quad \forall j.
\end{equation}
Note that $\Lambda_j$ now commutes with any charge degree of freedom.
These approximations are sufficient to continue our study.

The state~\eqref{Eq:tgge} satisfies Wick's theorem:
\begin{equation}
\langle c_{z}^{\dagger} c_{w}^{\dagger} c_{k} c_{q} \rangle_{\tau} = \langle c_{z}^{\dagger} c_{q} \rangle_{\tau} \langle c_{w}^{\dagger} c_{k} \rangle_{\tau} - \langle c_{z}^{\dagger} c_{k} \rangle_{\tau}\langle c_{w}^{\dagger} c_{q} \rangle_{\tau}
\end{equation}
and factorization in momentum space: $\langle c_{z}^{\dagger} c_{q} \rangle_{\tau} = \delta_{z,q}n_{q}(\tau)$, so that:
\begin{subequations}
\begin{align}
& \langle c_{z}^{\dagger} c_{w}^{\dagger} c_{k} c_{q} \rangle_{\tau} =  \left(\delta_{z,q} \delta_{w,k} -  \delta_{z,k}  \delta_{w,q} \right) n_{q}(\tau) n_{k}(\tau); \\
& \langle c_{z}^{\dagger} c_{w}^{\dagger} c_{k} c_{q} \rangle_{\tau} - \langle c_{q}^{\dagger} c_{k}^{\dagger} c_{w} c_{z} \rangle_{\tau} = 0.
\end{align}
\end{subequations}
Within the t-GGE approximation 
Starting from the following formula:
\begin{equation}
\begin{split}
n_{k} L'_j = & \Lambda_j  n_{k} \sqrt{\frac{\Gamma}{2}} \frac{1}{L} \sum_{w,q} e^{i (q+w) j} 2 \cos{(w)} a_{w} a_{q} = \\
= & \Lambda_j  \sqrt{\frac{\Gamma}{2}} \frac{1}{L} \sum_{w,q} e^{i (q+w) j} 2 \cos{(w)} a_{w} a_{q} \left(n_{k} - \delta_{k,w} - \delta_{k,q} \right),
\end{split}
\end{equation}
we obtain:

\begin{equation}
\left[ n_{k}, L'_j \right] =  - \sqrt{\frac{\Gamma}{2}} \frac{2}{L} \Lambda_j \sum_{q} e^{i (q+k) j} \left( \cos{(k)} - \cos{(k)} \right) a_{k} a_{q} ; 
\end{equation}
from which:
\begin{widetext}
\begin{equation}
L_j^{\prime \dagger} \left[ n_{k}, L'_j \right] =  -\frac{2 \Gamma}{L^2} \Lambda_j^\dagger \Lambda_j \sum_{q,w,z} e^{i (q+k) j} \left( \cos{(k)} - \cos{(k)} \right) \cos{(w)} a_z^{\dagger} a_{w}^{\dagger} a_{k} a_{q}.
\end{equation}
If we now sum over $j$ we are left with an expression where spin and charge are well separated:
\begin{equation}
\sum_j L_j^{\prime \dagger} \left[ n_{k}, L'_j \right] =  -\frac{2 \Gamma}{L^2} \left(  \frac{1}{N} \sum_{\ell}\Pi_{\ell,\ell+1} \right)  \sum_{q,w,z}\delta_{k+q, w+z} \left( \cos{(k)} - \cos{(k)} \right) \cos{(w)} \, a_z^{\dagger} a_{w}^{\dagger} a_{k} a_{q} .
\end{equation}
\end{widetext}
Moving to expectation values, we get:
\begin{align}
\frac{d}{d \tau} n_{k}(\tau) = -\frac{2 \Gamma}{L} \Pi(\tau) \sum_{q} (\cos{(k)}-\cos{(q)})^2 n_{q} (\tau) n_{k} (\tau) = 
\nonumber \\
=- 2 \Gamma \, \Pi(\tau) \int_{-\pi}^{+ \pi} \frac{dq}{2 \pi} (\cos{(k)}-\cos{(q)})^2 n_{q} (\tau) n_{k} (\tau).
\label{Eq.Rate.Equations}
\end{align}

Yet, at a more careful analysis, one finds that it is possible to give a better description of the dynamics by mixing the obtained rate equations with a mean-field behaviour:
\begin{widetext}
\begin{equation}
\frac{d}{d \tau} n_{k}(\tau) = - 2 \Gamma \, \Pi(\tau) \int_{-\pi}^{+ \pi} \frac{dq}{2 \pi} \left[ (1-\delta_{ \Psi_0}(\tau))+\delta_{ \Psi_0}(\tau) (\cos{(k)}-\cos{(q)})^2 \right] n_{q}  (\tau) n_{k} (\tau).
\label{Eq:re:delta}
\end{equation}
\end{widetext}

We verified the theoretical predictions given by Eq.~\eqref{Eq:re:delta} with exact numerical simulations of the effective master equation~\eqref{eq:eff_me} using the stochastic quantum trajectories approach; we have used the python-based QuTiP package~\cite{Qutip01, Qutip02} that allowed us to push our analysis up to $L = 12$ sites with high statistics ($N_{\rm traj} \geq 10^3$, $N_{\rm traj}$ being the number of trajectories).

\section{Evaluation of parameter $\delta_{\Psi_0}$ for the infinite-temperature state}
\label{App:delta:id}
{
Let us now discuss how to compute the parameter $\delta_{ \Psi_0}=1/8$ for the initial state that is an incoherent superposition of all spin states, also called the infinite-temperature state. We consider a single jump operator:
\begin{equation}
 L'_j = \sqrt{\frac{\Gamma}{2}} \left( 
 f_{j, \uparrow} f_{j+1,\downarrow} - f_{j,\downarrow} f_{j+1,\uparrow} +
 f_{j, \uparrow} f_{j-1,\downarrow} - f_{j,\downarrow} f_{j-1,\uparrow} 
 \right)
 \label{Eq:Jump:delta18}
\end{equation}
which acts on the three sites $j-1$, $j$ and $j+1$. 
There are 8 possible spin configurations on the three sites, and in a spin incoherent state all of them are possible with equal probability.
The jump operator~\eqref{Eq:Jump:delta18} has a different action on each of them:
\begin{subequations}
\begin{align}
 L'_j \ket{\uparrow \uparrow \uparrow} = & 0,
 \\
 L'_j \ket{\uparrow \uparrow \downarrow} = & - \sqrt{\frac \Gamma 2} \ket{\uparrow \circ \circ}, \\
 L'_j \ket{\uparrow \downarrow \uparrow} = & + \sqrt{\frac \Gamma 2} \ket{\uparrow \circ \circ}- \sqrt{\frac \Gamma 2} \ket{ \circ \circ \uparrow}, \label{Eq:Jump:Corr:1}
 \\
 L'_j \ket{\uparrow \downarrow \downarrow} = & -\sqrt{\frac \Gamma 2} \ket{ \circ \circ \downarrow}, \\
 L'_j \ket{\downarrow \uparrow \uparrow} = & +\sqrt{\frac \Gamma 2} \ket{ \circ \circ \uparrow},
 \\
 L'_j \ket{\downarrow \uparrow \downarrow} = & - \sqrt{\frac \Gamma 2} \ket{\downarrow \circ \circ}+ \sqrt{\frac \Gamma 2} \ket{ \circ \circ \downarrow}, \label{Eq:Jump:Corr:2} \\
 L'_j \ket{\downarrow \downarrow \uparrow} = & + \sqrt{\frac \Gamma 2} \ket{\downarrow \circ \circ}, \\
 L'_j \ket{\downarrow \downarrow \downarrow} = & 0.
\end{align}
\end{subequations}
In two situations, those of Eqs.~\eqref{Eq:Jump:Corr:1} and~\eqref{Eq:Jump:Corr:2}, the outcome state of the loss process has developed spatial quantum correlations, similarly to what was discussed for the N\'eel state in Eq.~\eqref{Eq:Neel:Jump:Correlated}.
In the other six situations, the outcome state does not feature any spatial quantum correlations.
In summary, the probability that a spin $\uparrow$ particle develops spatial quantum correlations is $1/8$; the same is true for a spin $\downarrow$ particle. This implies the use of $\delta_{ T=\infty}(\tau=0) = 1/8$ for the dissipative rate equations for an initial state that is a fully-incoherent spin mixture. \\

We conclude by remarking that in general the parameter can be expressed as the expectation value $\delta_{\Psi_{0}} =     \langle  \hat{\delta} \rangle_{\Psi_{0}}$ of a string of spin projectors, namely, given $P^{\uparrow}_j,P^{\downarrow}_j$ the projectors on spin up/down on the fermion $j$, the operator
\begin{equation}
\hat{\delta}=\frac{1}{2N} \sum_{j} \Big[ P^{\uparrow}_j  P^{\downarrow}_{j+1}   P^{\uparrow}_{j+2} +  P^{\downarrow}_j  P^{\uparrow}_{j+1}   P^{\downarrow}_{j+2} \Big].
\end{equation}
This expression can then be used to time evolve the parameter under non-hermitian spin evolution, as presented in the next section.

}

\section{Non-hermitian spin dynamics: time-dependent spin temperature $\beta_s(\tau)$ and time evolution of $\delta_{\Psi_0}$}
\label{App:solution:re}
\begin{figure}[t]
\includegraphics[width=\columnwidth] {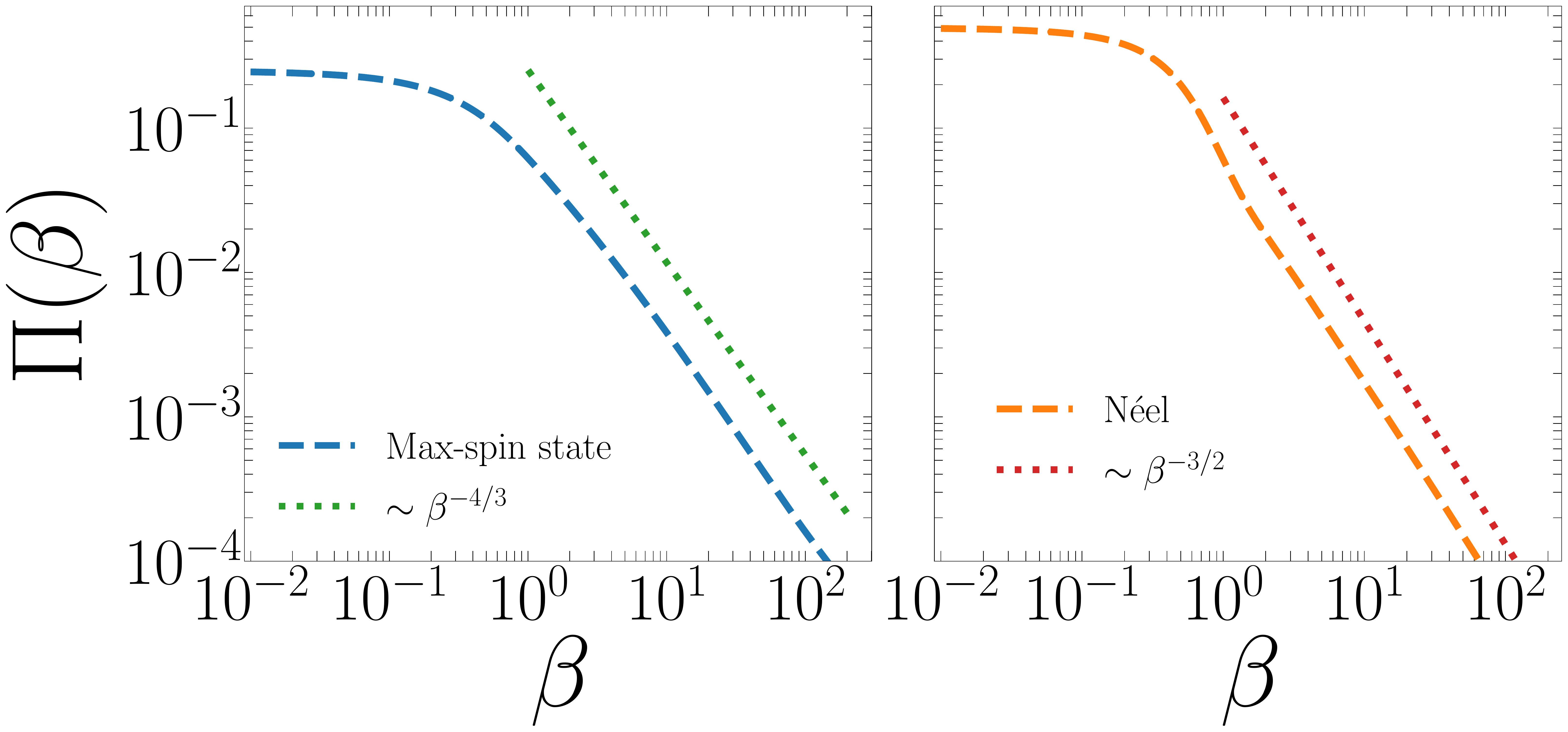}
 \caption{Non-Hermitian evolution for the spin-singlet projection operator for a (left) maximally-mixed spin state and (right) an ordered N\'eel state. Dashed lines: MPS data of the non-Hermitian Heisenberg evolution. Dotted lines: long-time behaviour. }
 \label{Fig:pi:dmrg}
\end{figure}

In order to solve the rate equations for the charge in~\eqref{Eq.Rate.Equations} that is Eq.~(9) in the main text while solving self-consistently the spin dynamics, we need to consider the spin dynamics.
The non-Hermitian spin
dynamics is governed by a ferromagnetic Heisenberg Hamitlonian, namely:
\begin{equation}
H_{s} = - \frac{\Gamma}{2} \sum_\ell \vec{\Sigma_\ell} \cdot\vec{\Sigma}_{\ell+1}; \qquad \Gamma >0.
\end{equation}
Within this framework~\cite{Sergi_2015}, we propose the following ansatz for the density matrix for the spin degrees of freedom:
\begin{equation}
\rho_{\mathrm{s}} (\tau) = \frac{e^{-\beta_s (\tau) H_{s}} \hspace{0.2cm} \rho(0) \hspace{0.2cm} e^{-\beta_s (\tau) H_{s}} }{\mathrm{Tr} \left[ e^{-2\beta_s (\tau) H_{s}} \hspace{0.2cm} \rho(0) \hspace{0.2cm}  \right]},
\label{Eq:rho:spin:supp}
\end{equation}
where $\rho(0)$ is the initial spin state, and in the article we have explicitly considered $\rho(0) = \mathbf{1} / d$ (with $d$ a normalization constant) for the maximally-mixed spin state, and $\rho(0) = \ket{\Psi_{\mathrm{Neel}}} \bra{\Psi_{\mathrm{Neel}}} $ for the ordered N\'eel state. 

\begin{figure}[t]
\includegraphics[width=\columnwidth] {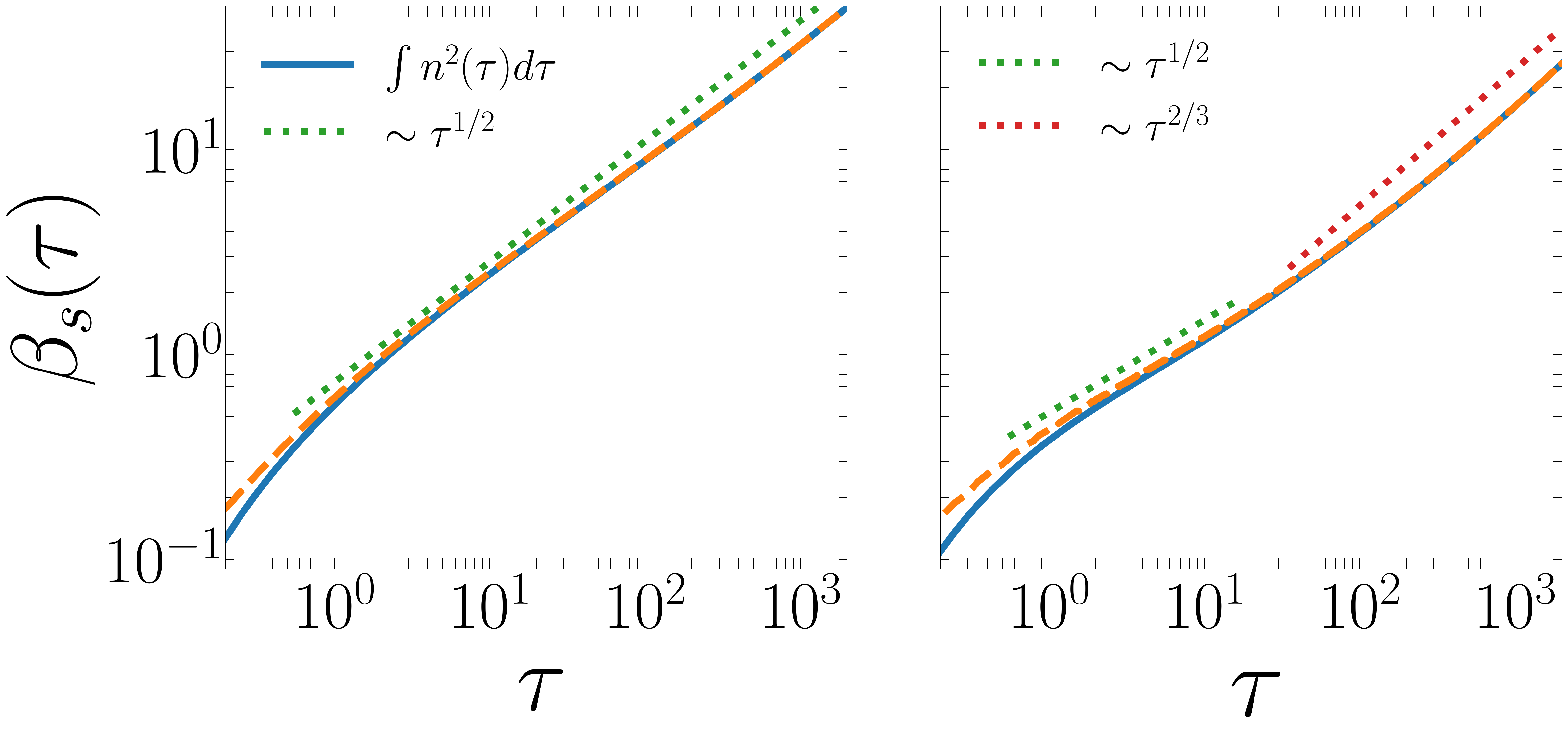}
 \caption{Time evolution of the effective spin chemical potential $\beta_s (\tau)$ (orange-dashed lines) compared with the time integral of the squared density (blue solid lines) for a maximally-mixed spin state (left) and an ordered N\'eel state (right). Dotted lines represent the theoretical trends. }
 \label{Fig:beta:tau}
\end{figure}

We employ matrix-product-states based algorithm~\cite{Schollwck2011} using the package ITensor~\cite{itensor} to reconstruct the expectation value of the operator 
\begin{equation}
 \Pi = \frac{1}{N} \sum_{\ell}\Pi_{\ell,\ell+1}
\end{equation}
for every value of $\beta_s$, so that the function $\Pi(\beta)$ is obtained; two examples are given in Fig.~\eqref{Fig:pi:dmrg}. 

\begin{figure}[t]
\includegraphics[width=0.8\columnwidth] {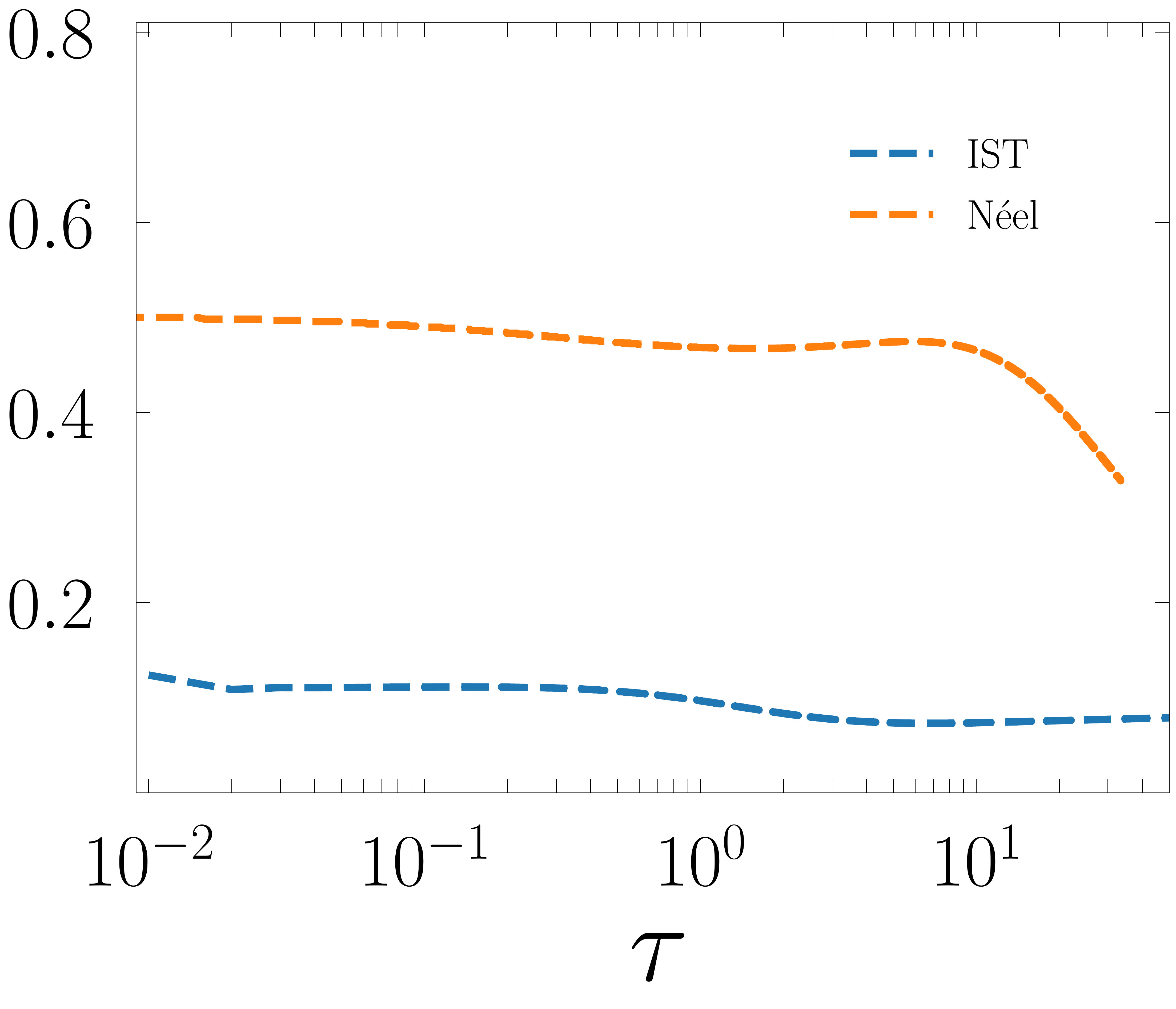}
 \caption{Time evolution of the parameter $\delta_{ \Psi_0} (\tau)$ for the maximally-mixed state (blue) and the N\'eel state (red). }
 \label{Fig:delta:tau}
\end{figure}

The dynamics of $\beta_s(\tau)$ is given by:
\begin{align}
\beta_s (\tau) =& \frac{1}{L} \sum_j \int_0^{\tau} \left \langle n_j n_{j+1} \right \rangle (\tau') \hspace{0.2cm} d \tau'= \nonumber \\=& \int_0^t d\tau' \int_{-\pi}^{\pi}\frac{dq }{2 \pi}   \int_{-\pi}^{\pi} \frac{d k}{2 \pi} n_k n_q (1 - \cos(q- k)) .
\label{Eq:beta:tau}
\end{align}
The specific value of $\Pi$ that should be taken at time $\tau$ thus requires the knowledge of the function $\beta_s(\tau)$. This turns the the rate equations into a set of integro-differential equations, for which we propose a numerical approximate solution.
We implement a $4^{\text{th}}$-order Runge-Kutta (RK) algorithm that assumes that $\Pi(\beta)$ is constant during the four intermediate steps of the RK algorithm. 
First, at fixed $\tau$ we compute $\beta_{\mathrm{s}}(\tau)$ according to Eq.~\eqref{Eq:beta:tau}, next we use a simple linear interpolation of the MPS data to obtain the value for $\Pi(\tau)$. Once the latter value has been obtained, we run the $4$ standard intermediate steps of the RK algorithm. We employ an integration step of $\mathrm{d} \tau = 10^{-2}$ and $N_{\mathrm{steps}} = 5 \cdot 10^4$, while we have discretised the $k$ space in $10^2$ points in between the range $\left[0, 2 \pi \right]$, which corresponds to consider a lattice with $L = 100$ sites. We conclude by displaying the data In Fig.~\ref{Fig:beta:tau} we display the data for the evolution of $\beta_s$ as a function of time.

The parameter $\delta_{ \Psi_0} (\tau)$ also depends on time, according to the same prescription of $\beta_s (\tau)$. We have investigated numerically such time evolution following the same procedure abovementioned. In Fig.~\ref{Fig:delta:tau} we show the results. We notice that $\delta_{ \Psi_0} (\tau)$ is slowly varying in the region of interest, and thus it can be approximated with the value at inital time. Moreover, for the infinite temperature state this value remains negligible in the time interval of interest justifying the mean-field description presented in Sec.~\ref{Sec:id:mf} of the main text for this initial state.

\bibliography{TwoBodyLossesHubbard.bib}
\end{document}